\documentclass[aps,prb,reprint,amsmath,amssymb,showpacs]{revtex4-1}

\usepackage{graphicx}
\usepackage{dcolumn}
\usepackage{bm}

\begin{document}

\title{Ferromagnetism and Fermi-surface transition
  in the periodic Anderson model:
Second-order phase transition without symmetry breaking}

\author{Katsunori Kubo}
\affiliation{
Advanced Science Research Center, Japan Atomic Energy Agency,
Tokai, Ibaraki 319-1195, Japan}

\date{\today}

\begin{abstract}
  We study ferromagnetism in the periodic Anderson model
  with and without a magnetic field by the Gutzwiller theory.
  We find three ferromagnetic phases:
  a weak ferromagnetic phase (FM0),
  a half-metallic phase without Fermi surface for the majority spin (FM1),
  and a ferromagnetic phase
  with almost completely polarized $f$-electrons (FM2).
  The Fermi surface changes
  from the large Fermi-surface in the paramagnetic state
  to the small Fermi-surface in FM2.
  We also find that the transitions between the ferromagnetic phases
  can be second-order phase transitions
  in spite of the absence of symmetry breaking.
  While we cannot define an order parameter for such transitions
  in an ordinary way,
  the topology of the Fermi surface characterizes the transitions,
  i.e., they are Lifshitz transitions.
\end{abstract}




\maketitle

\section{Introduction}
In heavy-fermion systems, external perturbations,
such as a magnetic field $H$ and pressure,
can change the electronic state drastically,
since the energy scale in the heavy-fermion systems is very low
due to the renormalization by the strong electron correlation.

The metamagnetic behavior
in CeRu$_2$Si$_2$~\cite{Besnus1985,Haen1987,Meulen1991,Aoki1993}
under a magnetic field
and the magnetic field induced transitions
in YbRh$_2$Si$_2$~\cite{Tokiwa2005,Rourke2008,Pfau2013,Pourret2013}
are typical examples of such effects.
At the transition field, the magnetization deviates substantially
from a linear dependence on $H$ observed
in lower fields.~\cite{Besnus1985,Haen1987,Tokiwa2005}
Such an anomaly in the magnetization indicates that
the electronic state is changed drastically at the transition.
Indeed, the effective mass deduced from the specific heat~\cite{Meulen1991}
and from the de Haas-van Alphen effect~\cite{Aoki1993} enhances
around the metamagnetic field in CeRu$_2$Si$_2$.
In YbRh$_2$Si$_2$, change in the Fermi surface
from the de Haas-van Alphen experiment~\cite{Rourke2008}
and anomaly in the thermoelectric power~\cite{Pfau2013,Pourret2013}
around 10~T have been reported.

Another examples are magnetic transitions and superconductivity under pressure,
such as ferromagnetic transitions~\cite{Pfleiderer2002}
and superconductivity~\cite{Saxena2000} in UGe$_2$.
There are two ferromagnetic phases in UGe$_2$:
the strongly polarized ferromagnetic phase phase under low pressure, called FM2,
and the ferromagnetic phase under high pressure, called FM1.
Under higher pressures, UGe$_2$ becomes paramagnetic.
The superconducting transition temperature becomes maximum around the pressure
where the FM1-FM2 transition temperature becomes zero.~\cite{Tateiwa2001}
The coefficient $A$ of $T^2$ term in the electrical resistivity,
$\sqrt{A}$ is proportional to the effective mass,
enhances in FM1.~\cite{Oomi1998,Saxena2000,Tateiwa2001,Settai2002}
The de Haas-van Alphen experiments show that the Fermi surface changes
at the ferromagnetic
transitions.~\cite{Terashima2001,Settai2002,Terashima2002,Haga2002,Settai2003}
These observations indicate that the electronic state changes drastically
at the ferromagnetic transitions.

To understand such phenomena,
we need a theory which
can describe the heavy-fermion state and the magnetically polarized state,
and can evaluate physical quantities
which reflect the change in the electronic state,
such as the effective mass.
To describe the heavy-fermion state,
the periodic Anderson model has been employed as a typical model.
While several approximations have been applied to the model,
the Gutzwiller method is a useful approximation
and succeeded in describing the heavy-fermion state.~\cite{Rice1986,Fazekas1987}
Thus, it is natural to extend the Gutzwiller method
for the model with magnetic polarization.
In fact, a similar approximation, that is,
the slave-boson mean-field theory of the Kotliar-Ruckenstein type,
has been applied to study the magnetization
of the model.~\cite{Reynolds1992,Dorin1993JAP,Dorin1993PRB}
However, the effects of magnetism and a magnetic field on the effective mass
have not been explored by these studies.

In this study, we extend the Gutzwiller method
for the magnetically polarized states,
and investigate ferromagnetic states at zero temperature.
We evaluate the magnetization and the effective mass.
We also investigate the Fermi-surface change
by ferromagnetism and a magnetic field.
Preliminary results on the magnetic field effect
have been reported in Ref.~\onlinecite{Kubo2013}.

This paper is organized as follows.
In Sec.~\ref{model},
we explain the periodic Anderson model.
In Sec.~\ref{method},
we introduce the variational wave function and the Gutzwiller approximation.
In Sec.~\ref{results},
we show the calculated results of physical quantities and phase diagrams.
We also discuss Fermi-surface states and the order of the phase transitions.
In Sec.~\ref{AF},
we discuss the antiferromagnetic states of the model.
In Sec.~\ref{summary},
we summarize the paper.

\section{Model}\label{model}
The periodic Anderson model is given by
\begin{equation}
  \begin{split}
    \mathcal{H}=&\sum_{\mathbf{k} \sigma}\epsilon_{\mathbf{k}}
    c^{\dagger}_{\mathbf{k} \sigma}c_{\mathbf{k} \sigma}
    +\sum_{i \sigma}\epsilon_f n_{f i \sigma}\\
    -&V\sum_{\mathbf{k} \sigma}(f^{\dagger}_{\mathbf{k} \sigma}c_{\mathbf{k} \sigma}
                            +c^{\dagger}_{\mathbf{k} \sigma}f_{\mathbf{k} \sigma})
    +U\sum_{i}n_{f i \uparrow}n_{f i \downarrow},
  \end{split}
\end{equation}
where
$c^{\dagger}_{\mathbf{k} \sigma}$
and
$f^{\dagger}_{\mathbf{k} \sigma}$
are the creation operators of the conduction and $f$ electrons,
respectively, with momentum $\mathbf{k}$ and spin $\sigma$.
$n_{f i \sigma}$ is the number operator
of the $f$ electron with spin $\sigma$ at site $i$.
$\epsilon_{\mathbf{k}}$ is the kinetic energy of the conduction electron,
$\epsilon_f$ is the $f$-electron level,
$V$ is the hybridization matrix element,
and $U$ is the onsite Coulomb interaction between $f$ electrons.
The spatial extent of the $f$-electron wave-function is narrow and
the Coulomb interaction between $f$ electrons is large,
and thus, we set $U \rightarrow \infty$ for simplicity.
We set the energy level of the conduction electrons as the
origin of energy, that is, $\sum_{\mathbf{k}}\epsilon_{\mathbf{k}}=0$.

Under a finite magnetic field $H$,
we replace $\epsilon_f$ by $\epsilon_{f \sigma}=\epsilon_f-\sigma H$,
where $\sigma=+1$ ($-1$) for $\uparrow$ ($\downarrow$) spin
in the right hand side of the equation.
Here, we have set the Bohr magneton $\mu_{\text{B}}=1$
as the unit of magnetization.
We have set the $g$-factors $g_f=2$ for $f$ electrons
and $g_c=0$ for conduction electrons, that is,
we ignore the Zeeman term for the conduction electrons.
As we will show later, the polarization of the conduction electrons
is small even in ferromagnetic phases,
and this assumption is justified as long as the magnetic field
is not very large.

Experimentally, magnetic anisotropy is important in $f$-electron systems,
while it is not included in the present model.
To interpret experimental results,
we should regard the magnetization and magnetic field of the present theory
as being along the easy axis of the materials.

\section{Method}\label{method}
In this study, we focus on ferromagnetism and
magnetic field effects on the paramagnetic state,
and then, we assume a spatially uniform state.
The variational wave function is given by
\begin{equation}
  | \psi \rangle
  =P | \phi_{\uparrow} \rangle \otimes | \phi_{\downarrow} \rangle,
\end{equation}
where
$P=\prod_{i}[1-n_{f i \uparrow}n_{f i \downarrow}]$
excludes the double occupancy of the $f$ electrons at the same site.
For the one-electron part of the wave function,
we consider the following form:
\begin{equation}
  | \phi_{\sigma} \rangle=
  \prod_{k<k_{\mathrm{F} \sigma}}
  [c^{\dagger}_{\mathbf{k} \sigma}
  +a_{\sigma}(\mathbf{k})f^{\dagger}_{\mathbf{k} \sigma}]
  | 0\rangle,
\end{equation}
for $n_{\sigma} < 1$,
where $n_{\sigma}$ is the total number of the spin-$\sigma$ electrons per site
and $k_{\mathrm{F} \sigma}$ is the Fermi momentum for spin $\sigma$.
$a_{\sigma}(\mathbf{k})$ are spin-dependent variational parameters.
For $n_{\sigma} > 1$,
both the hybridized bands are filled below $k_{\mathrm{F} \sigma}$
and only the lower hybridized band is filled above $k_{\mathrm{F} \sigma}$
for $U=0$,
and thus, we consider the one-electron part given by
\begin{equation}
  | \phi_{\sigma} \rangle=
  \prod_{p<k_{\mathrm{F} \sigma}}
  c^{\dagger}_{\mathbf{p} \sigma}
  f^{\dagger}_{\mathbf{p} \sigma}
  \prod_{k>k_{\mathrm{F} \sigma}}
  [c^{\dagger}_{\mathbf{k} \sigma}
  +a_{\sigma}(\mathbf{k})f^{\dagger}_{\mathbf{k} \sigma}]
  | 0\rangle.
\end{equation}

By using the Gutzwiller
approximation,~\cite{Fazekas1987,Kubo2011JPSJ80.06,Kubo2011JPSJ80.11}
we evaluate the expectation values of physical quantities
of the variational wave function.
%
The $f$-electron number $n_{f \sigma}$ per site with spin $\sigma$
is given by
\begin{equation}
  n_{f \sigma}
  =\frac{1}{L}\sum_{k<k_{\text{F} \sigma}}
  \frac{a^2_{\sigma}(\mathbf{k})}{q^{-1}_{\sigma}+a^2_{\sigma}(\mathbf{k})},
  \label{eq:nf1}
\end{equation}
for $n_{\sigma} < 1$,
and
\begin{equation}
  n_{f \sigma}
  =\frac{1}{L}\sum_{k>k_{\text{F} \sigma}}
  \frac{a^2_{\sigma}(\mathbf{k})}{q^{-1}_{\sigma}+a^2_{\sigma}(\mathbf{k})}
  +n_{\sigma}-1,
  \label{eq:nf2}
\end{equation}
for $n_{\sigma} > 1$,
where $L$ is the number of the lattice sites and
\begin{equation}
  q_{\sigma}=\frac{1-n_f}{1-n_{f \sigma}},
\end{equation}
with $n_f=\sum_{\sigma}n_{f \sigma}$.
$n_{\sigma}-1$ in Eq.~\eqref{eq:nf2} is the $f$-electron number
with spin $\sigma$ inside the Fermi momentum $k_{\text{F} \sigma}$.

We evaluate the momentum distribution functions
$n_{c \sigma}(\mathbf{k})
=\langle c^{\dagger}_{\mathbf{k} \sigma} c_{\mathbf{k} \sigma} \rangle
=\langle \psi |c^{\dagger}_{\mathbf{k} \sigma} c_{\mathbf{k} \sigma} | \psi \rangle
/\langle \psi | \psi \rangle$
of the conduction electrons and
$n_{f \sigma}(\mathbf{k})
=\langle f^{\dagger}_{\mathbf{k} \sigma} f_{\mathbf{k} \sigma} \rangle$
of the $f$ electrons.
For $n_{\sigma}<1$, we obtain
\begin{equation}
  n_{c \sigma}(\mathbf{k})
  =
  \begin{cases}
    \Delta n_{c \sigma}(\mathbf{k})
    & \text{for $k<k_{\text{F} \sigma}$} \\
    0
    & \text{for $k>k_{\text{F} \sigma}$}
  \end{cases},
  \label{eq:nck1}
\end{equation}
with
\begin{equation}
  \Delta n_{c \sigma}(\mathbf{k})
  =\frac{q^{-1}_{\sigma}}{q^{-1}_{\sigma}+a^2_{\sigma}(\mathbf{k})},
  \label{eq:Dnc1}
\end{equation}
and
\begin{equation}
  n_{f \sigma}(\mathbf{k})
  =
  \begin{cases}
    (1-q_{\sigma})n_{f \sigma}+\Delta n_{f \sigma}(\mathbf{k})
    & \text{for $k<k_{\text{F} \sigma}$} \\
    (1-q_{\sigma})n_{f \sigma}
    & \text{for $k>k_{\text{F} \sigma}$}
  \end{cases},
  \label{eq:nfk1}
\end{equation}
with
\begin{equation}
  \Delta n_{f \sigma}(\mathbf{k})
  =q_{\sigma}
  \frac{a^2_{\sigma}(\mathbf{k})}{q^{-1}_{\sigma}+a^2_{\sigma}(\mathbf{k})}.
  \label{eq:Dnf1}  
\end{equation}
For $n_{\sigma}>1$, we obtain
\begin{equation}
  n_{c \sigma}(\mathbf{k})
  =
  \begin{cases}
    1
    & \text{for $k<k_{\text{F} \sigma}$} \\
    1-\Delta n_{c \sigma}(\mathbf{k})
    & \text{for $k>k_{\text{F} \sigma}$}
  \end{cases},
  \label{eq:nck2}
\end{equation}
with
\begin{equation}
  \Delta n_{c \sigma}(\mathbf{k})
  =\frac{a^2_{\sigma}(\mathbf{k})}{q^{-1}_{\sigma}+a^2_{\sigma}(\mathbf{k})},
  \label{eq:Dnc2}
\end{equation}
and
\begin{equation}
  n_{f \sigma}(\mathbf{k})
  =
  \begin{cases}
    q_{\sigma}+(1-q_{\sigma})n_{f \sigma}
    & \text{for $k<k_{\text{F} \sigma}$} \\
    q_{\sigma}+(1-q_{\sigma})n_{f \sigma}-\Delta n_{f \sigma}(\mathbf{k})
    & \text{for $k>k_{\text{F} \sigma}$}
  \end{cases},
  \label{eq:nfk2}
\end{equation}
with
\begin{equation}
  \Delta n_{f \sigma}(\mathbf{k})
  =q_{\sigma}\frac{q^{-1}_{\sigma}}{q^{-1}_{\sigma}+a^2_{\sigma}(\mathbf{k})}.
  \label{eq:Dnf2}
\end{equation}

Energy per site is given by
\begin{equation}
  e=\frac{\langle \mathcal{H} \rangle}{L} =\sum_{\sigma}e_{\sigma},
\end{equation}
where
\begin{equation}
  e_{\sigma}
  =\frac{1}{L}\sum_{k<k_{\text{F} \sigma}}\epsilon_{\mathbf{k}}
  +\frac{1}{L}\sum_{k<k_{\text{F} \sigma}}
  \frac{(\epsilon_{f \sigma}-\epsilon_{\mathbf{k}})a^2_{\sigma}(\mathbf{k})
    -2Va_{\sigma}(\mathbf{k})}{q^{-1}_{\sigma}+a^2_{\sigma}(\mathbf{k})},
  \label{eq:E1}
\end{equation}
for $n_{\sigma}<1$,
and
\begin{equation}
  e_{\sigma}
  =\epsilon_{f \sigma}(n_{\sigma}-1)
  +\frac{1}{L}\sum_{k>k_{\text{F} \sigma}}
  \frac{(\epsilon_{f \sigma}-\epsilon_{\mathbf{k}})a^2_{\sigma}(\mathbf{k})
    -2Va_{\sigma}(\mathbf{k})}{q^{-1}_{\sigma}+a^2_{\sigma}(\mathbf{k})},
  \label{eq:E2}
\end{equation}
for $n_{\sigma}>1$.

Now, we minimize the energy with respect to the variational parameters
$a_{\sigma}(\mathbf{k})$.
From $\partial e/ \partial a_{\sigma}(\mathbf{k})=0$, we obtain
\begin{equation}
  a_{\sigma}(\mathbf{k})
  =\frac{2V}{\tilde{\epsilon}_{f \sigma}-\epsilon_{\mathbf{k}}
    +\sqrt{(\tilde{\epsilon}_{f \sigma}-\epsilon_{\mathbf{k}})^2
      +4\tilde{V}^2_{\sigma}}},
  \label{eq:a}
\end{equation}
where $\tilde{V}_{\sigma}=\sqrt{q_{\sigma}} V$
and $\tilde{\epsilon}_{f \sigma}$ is the renormalized $f$-level.
The renormalized $f$-level $\tilde{\epsilon}_{f \sigma}$ satisfy
\begin{equation}
  \epsilon_{f \sigma}-\tilde{\epsilon}_{f \sigma}=
  -\sum_{\sigma^{\prime}}
  \tilde{V}^2_{\sigma^{\prime}} I_{2 \sigma^{\prime}}
  q_{\sigma^{\prime}}
  \frac{\partial q^{-1}_{\sigma^{\prime}}}{\partial n_{f \sigma}}. \label{eq:ef}
\end{equation}
The integral is defined by
\begin{equation}
  I_{l \sigma}=
  \frac{1}{L}{\sum_{\mathbf{k}}}^{\prime}
  \frac{(\epsilon_{\mathbf{k}}-\tilde{\epsilon}_{f \sigma})^{l-2}}
  {\sqrt{(\epsilon_{\mathbf{k}}-\tilde{\epsilon}_{f \sigma})^2
      +4\tilde{V}^2_{\sigma}}},
  \label{eq:I}
\end{equation}
where ${\sum_{\mathbf{k}}}^{\prime}$ means that the summation runs over
$k<k_{\text{F} \sigma}$ for $n_{\sigma}<1$
and
$k>k_{\text{F} \sigma}$ for $n_{\sigma}>1$.
We can rewrite Eqs.~\eqref{eq:nf1} and \eqref{eq:nf2}
by using Eqs.~\eqref{eq:a} and \eqref{eq:I}:
\begin{equation}
  n_{f \sigma}=\frac{n_{\sigma}+I_{3 \sigma}}{2}. \label{eq:nf}
\end{equation}

We solve Eqs. \eqref{eq:ef} and \eqref{eq:nf}
with respect to $\tilde{\epsilon}_{f \sigma}$ and $n_{f \sigma}$
for each value of the total polarization $M^{\prime}=n_{\uparrow}-n_{\downarrow}$
fixing the total number $n=n_{\uparrow}+n_{\downarrow}$ of electrons per site,
and evaluate the energy.
$n_{\sigma}$ can be tuned by varying the Fermi momentum $k_{\text{F} \sigma}$.
Then, we determine $M^{\prime}$ for which the energy is the lowest,
and evaluate physical quantities.
Here, we note that Eqs. \eqref{eq:ef} and \eqref{eq:nf}
can be derived also by the slave-boson mean-field theory
of the Kotliar-Ruckenstein type,~\cite{Dorin1993JAP}
and several physical quantities, such as magnetism, are equivalent
between the slave-boson mean-field theory and the Gutzwiller method.
However, some quantities are difficult to be determined within
the slave-boson mean-field theory.
For example, for $n_{\sigma}<1$, the electron distribution function is always
the Fermi distribution function, that is,
unity below the Fermi momentum and zero above the Fermi momentum,
since the slave-boson mean-field theory is a one-particle approximation.
Thus, we obtain
$\Delta n_{c \sigma}(k_{\text{F} \sigma})+\Delta n_{f \sigma}(k_{\text{F} \sigma})=1$.
On the other hand, we can deal with the renormalization effect
on the electron distribution by the Gutzwiller method
as shown in Eqs. \eqref{eq:nck1}--\eqref{eq:Dnf2}.

\section{Results}\label{results}
\begin{figure}
  \includegraphics[width=0.98\linewidth]
  {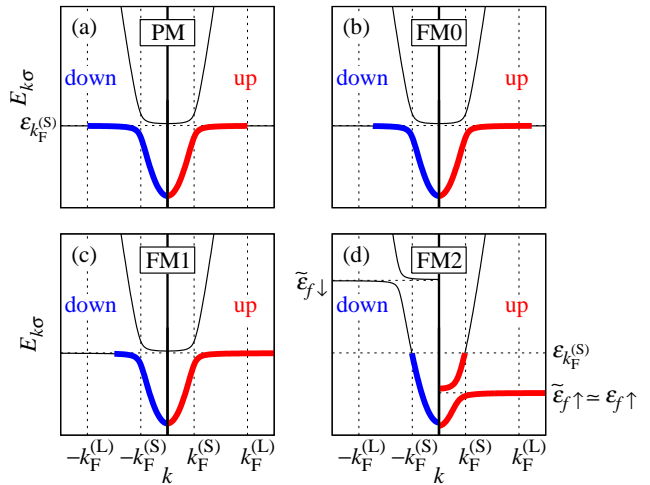}
  \caption{\label{bands}
    (Color online)
    Schematic band structures
    of paramagnetic and ferromagnetic phases in the periodic Anderson model.
    (a) paramagnetic phase (PM),
    (b) weakly polarized ferromagnetic phase (FM0),
    (c) half-metallic phase (FM1),
    and
    (d) ferromagnetic phase with an almost completely
    polarized $f$-electron state (FM2).
    $E_{k \sigma}$ denotes the quasi-particle energy.
    Right (left) part of each figure shows the up- (down-) spin band.
    The occupied states are represented by the bold lines.
  }
\end{figure}

Before presenting our calculated results,
we discuss possible ferromagnetic states of the model
by using schematic band structures shown in Fig.~\ref{bands}.

In the paramagnetic phase (PM), Fig.~\ref{bands}(a),
the numbers of up- and down-spin electrons are the same.
The effective $f$-level
$\tilde{\epsilon}_{f \uparrow}=\tilde{\epsilon}_{f \downarrow}$
is renormalized to a value around
the kinetic energy $\epsilon_{k^{\text{(S)}}_{\text{F}}}$
at the Fermi momentum for the small Fermi-surface
as long as $\epsilon_f \ll \epsilon_{k^{\text{(S)}}_{\text{F}}}$,
and the Fermi level is also near $\epsilon_{k^{\text{(S)}}_{\text{F}}}$.
Here, the small Fermi-surface is defined as the Fermi surface
for the state where the $f$ orbital with $n_f=1$
is assumed to be decoupled from the conduction electrons.
However, the $f$-electron state contributes to the band,
and the large Fermi-surface realizes
with the Fermi momentum $k^{\text{(L)}}_{\text{F}}$
which includes the $f$-electron contribution.
As a result,
the dispersion around Fermi momentum $k^{\text{(L)}}_{\text{F}}$ is weak,
and a heavy-electron state realizes.
Note that if $\epsilon_f \gg \epsilon_{k^{\text{(S)}}_{\text{F}}}$,
the renormalization is weak and
$\tilde{\epsilon}_{f \uparrow}=\tilde{\epsilon}_{f \downarrow} \simeq \epsilon_f$.

In a state with weak polarization,
by a spontaneous phase transition or by a magnetic field,
the band structure will become like Fig.~\ref{bands}(b).
Here, we call this state FM0.

When the polarization becomes larger,
the lower hybridized band will be filled up
by the up-spin electrons as shown in Fig.~\ref{bands}(c).
We call this state FM1.
In this state, the Fermi surface for the up-spin states disappears,
that is, this is a half-metallic state.
There is a hybridization gap,
and this state will be stable in some degree.
This half-metallic state has been obtained by
the slave-boson mean-field theory~\cite{Reynolds1992,Dorin1993JAP,Dorin1993PRB}
and in the Kondo lattice model.~\cite{Irkhin1991,Watanabe2000,
  Kusminskiy2008,Beach2008,Peters2012PRL,Bercx2012,Peters2012PRB}

When the polarization increases further,
the up-spin electrons start to fill the upper hybridized band
as in Fig.~\ref{bands}(d).
We call this state FM2.
In FM2, the $f$ electrons will polarize almost completely,
that is, $n_{f \uparrow} \simeq 1$ and $n_{f \downarrow} \simeq 0$.
Since $n_{f \downarrow} \simeq 0$,
the up-spin electrons can move almost freely,
and the effective $f$-level for up spin should be near the bare $f$-level.
On the other hand,
the down-spin electrons experience the Coulomb interaction strongly
for $n_{f \uparrow} \simeq 1$,
and the effective $f$-level for down spin becomes
much higher than the Fermi level.
As a result, the electronic state around the Fermi surface
is composed mostly of the conduction-electron states,
and the Fermi surface is similar to that expected
for the small Fermi-surface state.
Such a small Fermi-surface induced by
ferromagnetism and/or magnetic field has been discussed
in Refs.~\onlinecite{Miyake2006,Suzuki2010}.

Now, we show the calculated results in the following subsections.
In the present study, we consider a simple model for the conduction band.
The density of states of the conduction electrons
is given by $\rho(\epsilon)=1/(2W)$ for $|\epsilon|<W$
and $\rho(\epsilon)=0$ otherwise.
We expect that a change in the form of $\rho(\epsilon)$ will affect
the results little unless $\rho(\epsilon)$ has some characteristic structures,
such as strong peaks.

\subsection{Phase diagram}
\begin{figure}
  \includegraphics[width=0.95\linewidth]
  {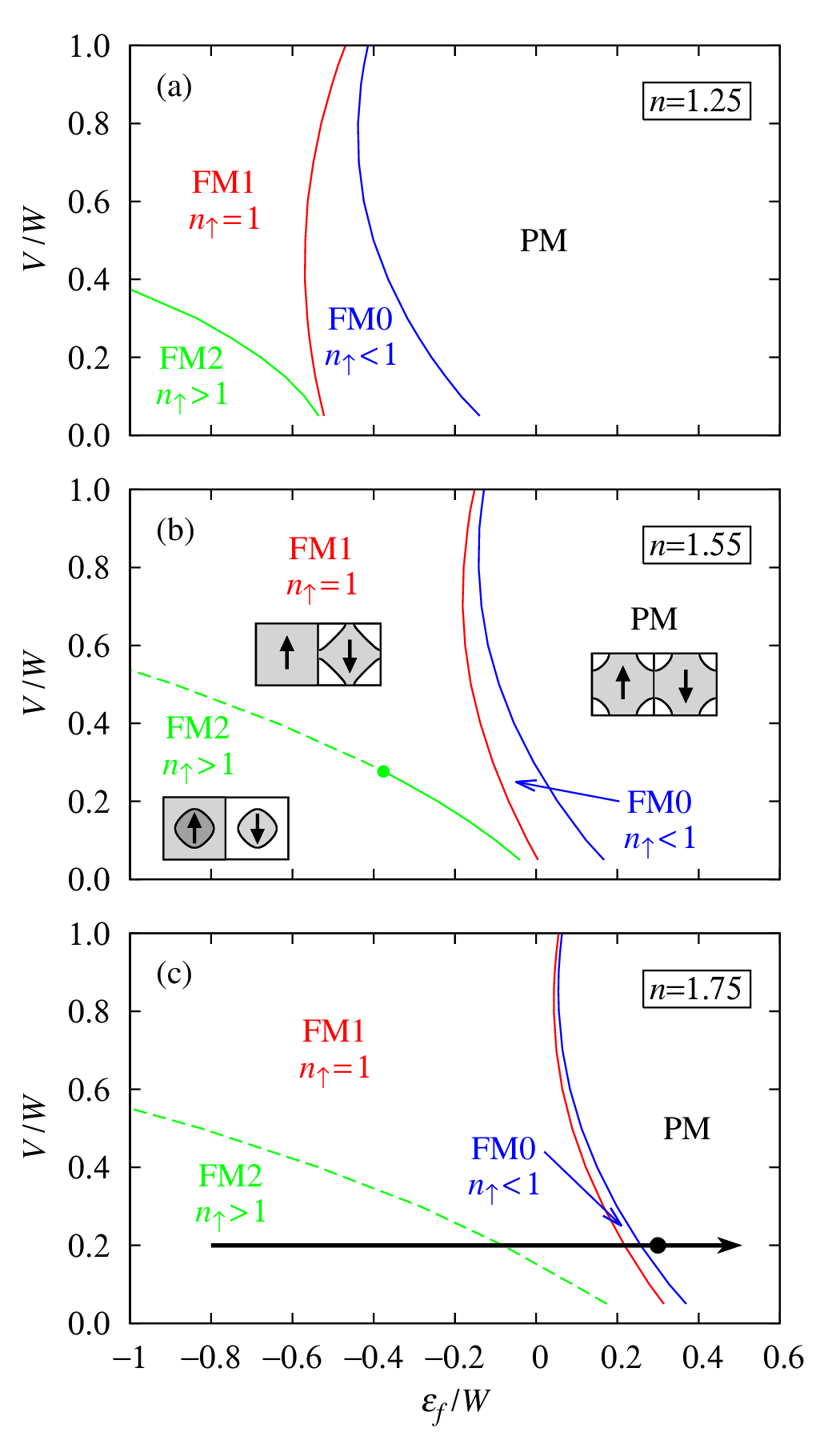}
  \caption{\label{PD}
    (Color online)
    Phase diagrams
    (a) for $n=1.25$,
    (b) for $n=1.55$,
    and
    (c) for $n=1.75$.
    The solid lines indicate second-order phase transitions
    and
    the dashed lines indicate first-order phase transitions.
    The circle in (b) represents the end point
    of the first-order phase transition.
    In (b), we also show the Fermi-surface structures
    in these phases schematically.
    The lower hybridized band is occupied by electrons
    in the lightly shaded area.
    In the darkly shaded area,
    both the lower and upper hybridized bands are filled by electrons.
    The bold arrow in (c) indicates
    the parameters for which
    physical quantities shown in Figs.~\ref{vary_ef_1} and \ref{vary_ef_2}
    are calculated.
    The circle in (c) indicates the parameter set for which
    the magnetic field dependences of the physical quantities
    are calculated (Figs.~\ref{vary_H_1} and \ref{vary_H_2}).
  }
\end{figure}

In Fig.~\ref{PD}, we show the phase diagrams
for the total number of electrons $n=1.25$, 1.55, and 1.75 per site.
All the three ferromagnetic phases discussed above, FM0, FM1, and FM2, appear.
For $n=1.25$, Fig.~\ref{PD}(a), all the phase transition are of second order.
On the other hand, the FM1-FM2 transition is of first order for $n=1.75$
[Fig.~\ref{PD}(c)].
Here, we discuss the reason why the order of the FM1-FM2 transition
changes with $n$.
As we will show later [Fig.~\ref{vary_ef_1}(a)],
the polarization of the conduction electrons
$M_c=n_{c \uparrow}-n_{c \downarrow}
=(n_{\uparrow}-n_{f \uparrow})-(n_{\downarrow}-n_{f \downarrow})$
is small even in the ferromagnetic phases.
Thus, the magnetization in FM1 ($n_{\uparrow}=1$) is approximated as
$M=n_{f \uparrow}-n_{f \downarrow} \simeq n_{\uparrow}-n_{\downarrow}
= 2n_{\uparrow}-n=2-n$,
and $M$ is smaller for larger $n$.
In FM2, $M$ is almost 1, irrespective of $n$.
Then, the change in $M$ at the FM1-FM2 transition is larger for large $n$.
Such a large change in the electronic state may not occur continuously
but tends to occur through a first order transition, as for $n=1.75$.
A similar discussion has been applied
for the valence transition in the periodic Anderson model
with an interorbital Coulomb interaction.~\cite{Kubo2011JPSJ80.06}
For an intermediate value of $n$,
we find the end point of the first order transition
as shown in Fig.~\ref{PD}(b).
On the second order transition line, the magnetic susceptibility diverges.
At the end point, the valence susceptibility $dn_f/d\epsilon_f$ also diverges.
Such fluctuations of various kinds may induce interesting phenomena,
e.g., unconventional superconductivity.

From these phase diagrams,
we gain an insight into the effects of pressure on Ce and Yb compounds.
$\epsilon_f$ describes the one-electron level for a Ce compound
and one-hole level for an Yb compound.
Then, $\epsilon_f$ will increase by pressure for a Ce compound,
but will decrease for an Yb compound,
since negatively charged ions surrounding a positively charged rare-earth ion
will become close to the rare-earth ion.
On the other hand,
$V$ and $W$ increase under a pressure irrespective of compounds.
Thus, it is not obvious whether
the effect of pressure is opposite or not between Ce and Yb compounds.
In the present model, there are two independent parameters
$\epsilon_f/W$ and $V/W$ except for the overall energy scale.
From the above phase diagrams,
we observe that we can change the electronic state easier
by varying $\epsilon_f/W$, e.g., along the bold arrow in Fig.~\ref{PD}(c),
than by varying $V/W$,
and we may ignore the change in $V/W$ under pressure as an approximation.

Here, we further assume that
the changes by a pressure $p$ are approximated linear in $p$.
Then, we can express $\epsilon_f=\epsilon^{(0)}_f+a_{\epsilon_f} p$
and $W=W^{(0)}+a_W p$,
where $\epsilon^{(0)}_f$ is the $f$ level at $p=0$,
$W^{(0)}$ is the band width at $p=0$,
$a_{\epsilon_f}>0$ for Ce compounds, $a_{\epsilon_f}<0$ for Yb compounds,
and $a_W>0$.
The ratio $\epsilon_f/W$ under pressure is given by
\begin{equation}
  \frac{\epsilon_f}{W}=\frac{\epsilon^{(0)}_f+\tilde{a}_{\epsilon_f} p}{W^{(0)}},
\end{equation}
with $\tilde{a}_{\epsilon_f}=a_{\epsilon_f}-a_W \epsilon^{(0)}_f/W^{(0)}$.
For Ce compounds with $\epsilon^{(0)}_f<0$,
typical for magnetically ordered materials at ambient pressure,
we obtain $\tilde{a}_{\epsilon_f}>0$.
For Yb compounds with $\epsilon^{(0)}_f>0$,
typical for paramagnetic materials at ambient pressure,
$\tilde{a}_{\epsilon_f}<0$.
Thus, magnetically ordered states of Ce compounds will be
destabilized by pressure,
and paramagnetic Yb compounds may become magnetic under pressure.
In the above sense, the pressure effects on Ce and Yb compounds are opposite.

However, the pressure effects on Ce compounds with $\epsilon^{(0)}_f>0$
and on Yb compounds with $\epsilon^{(0)}_f<0$ depend on the details of
the parameters.
Thus, in principle, paramagnetic Ce compounds can become magnetic
and magnetically ordered states of Yb compounds can become paramagnetic
under pressure, when the effect of pressure on the band width is large.

\subsection{$\epsilon_f$ dependence}
\begin{figure}
  \includegraphics[width=0.95\linewidth]
  {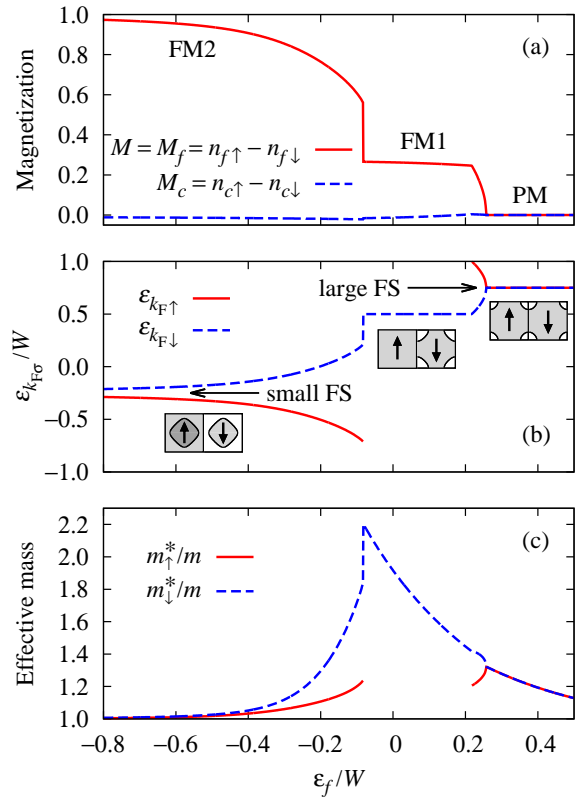}
  \caption{\label{vary_ef_1}
    (Color online)
    Physical quantities as functions of $\epsilon_f$
    for $n=1.75$ and $V/W=0.2$.
    (a) magnetization $M=M_f$ and polarization of the conduction band $M_c$,
    (b) kinetic energy $\epsilon_{k_{\text{F}\sigma}}$
    of the conduction electron at the Fermi momentum,
    and
    (c) effective mass $m^*_{\sigma}$.
    In (b), we show the Fermi-surface (FS) structure
    in each phase schematically.
  }
\end{figure}

Next, we show $\epsilon_f$ dependences of physical quantities.
In Fig.~\ref{vary_ef_1},
we show the magnetization,
the kinetic energy $\epsilon_{k_{\text{F} \sigma}}$ of the conduction electron
at the Fermi momentum,
and the effective mass for $n=1.75$ and $V/W=0.2$.
$\epsilon_f$ is varied along the bold arrow in Fig.~\ref{PD}(c).

The magnetization [Fig.~\ref{vary_ef_1}(a)] is almost 1 in FM2,
decreases by increasing $\epsilon_f$,
and the state changes to the FM1, FM0, and PM states.
Even in the ferromagnetic phases,
the polarization of the conduction electrons $M_c$ is small,
since the loss of the kinetic energy is large
for a large polarization of the conduction electrons.

In Fig.~\ref{vary_ef_1}(b), we show $\epsilon_{k_{\text{F} \sigma}}$.
In the present theory, physical quantities depend on momentum $\mathbf{k}$
only through $\epsilon_{\mathbf{k}}$,
and here we show $\epsilon_{k_{\text{F} \sigma}}$ instead of
the Fermi momentum $k_{\text{F} \sigma}$ itself.
If the explicit form of the dispersion $\epsilon_{\mathbf{k}}$ is given,
we can extract $k_{\text{F} \sigma}$ from $\epsilon_{k_{\text{F} \sigma}}$.
In FM2, $\epsilon_{k_{\text{F} \sigma}}$ has a value
around that for the small Fermi-surface state.
In FM1, the Fermi surface for the up-spin state disappears.
In PM, the large Fermi-surface state realizes.

Figure~\ref{vary_ef_1}(c) shows the effective mass.
For the periodic Anderson model,
the effective mass is usually defined
by the inverse of the renormalization factor for the $f$ electrons.
This is reasonable as long as the Fermi surface is composed
mainly of $f$ electrons as in a state with very large effective-mass.
However, in magnetically ordered states and in a state under a magnetic field,
the $f$-electron contribution to the Fermi surface can become small.
Thus, we should better to define the effective mass
by the renormalization of the hybridized band.
In this study, we define the spin-dependent effective-mass $m^*_{\sigma}$
by the jump in the momentum distribution at the Fermi momentum:
\begin{equation}
  \frac{m^*_{\sigma}}{m}
  =\frac{1}{\Delta n_{c \sigma}(k_{\text{F} \sigma})
           +\Delta n_{f \sigma}(k_{\text{F} \sigma})},
\end{equation}
where $m$ is the bare electron mass.

The effective mass in FM2 becomes small by decreasing $\epsilon_f$,
since the magnetization becomes large and the correlation effects become weaker.
In the PM phase, the number of $f$ electrons decreases
as $\epsilon_f$ increases,
and then, the correlation effects becomes less significant
and the effective mass decreases.
In between, in FM1, the effective mass for the down-spin electrons has a peak.
Note that we cannot define the effective mass for the up-spin state in FM1,
since there is no Fermi surface for the up-spin states.

\begin{figure}
  \includegraphics[width=0.95\linewidth]
  {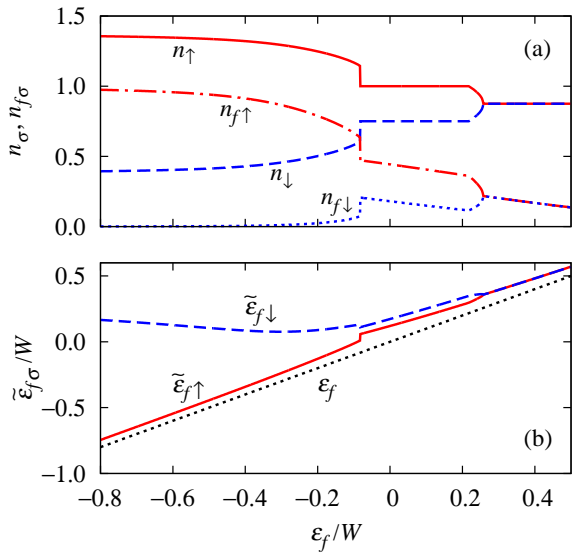}
  \caption{\label{vary_ef_2}
    (Color online)
    Physical quantities as functions of $\epsilon_f$
    for $n=1.75$ and $V/W=0.2$.
    (a) $n_{\sigma}$ and $n_{f\sigma}$
    and
    (b) $\tilde{\epsilon}_{f\sigma}$.
    The dotted line in (b) indicates the bare $f$-level $\epsilon_f$.
  }
\end{figure}

Figure~\ref{vary_ef_2}(a) shows
the number of electrons $n_{\sigma}$ with spin $\sigma$
and the number of $f$ electrons $n_{f \sigma}$ with spin $\sigma$.
In FM2, the polarization of $f$ electrons is almost complete,
that is, $n_{f \uparrow} \simeq 1$ and $n_{f \downarrow} \simeq 0$.
In FM1, i.e., the half-metallic state, $n_{\uparrow}=1$.

Figure~\ref{vary_ef_2}(b) shows the renormalized $f$-level.
In FM2, $\tilde{\epsilon}_{f \downarrow}$ is much larger than
$\epsilon_{k^{\text{(S)}}_{\text{F}}}=-0.25W$
and $\tilde{\epsilon}_{f \uparrow} \simeq \epsilon_f$.
In FM1, FM0, and PM,
$\epsilon_f$ is larger than $\epsilon_{k^{\text{(S)}}_{\text{F}}}$,
and the renormalization effect is weak, that is,
$\tilde{\epsilon}_{f \sigma} \simeq \epsilon_f$.

The overall behaviours of the magnetization [Fig.~\ref{vary_ef_1}(a)]
and the effective mass [Fig.~\ref{vary_ef_1}(c)]
as functions of $\epsilon_f$
are similar to those as functions of pressure in UGe$_2$.~\cite{Oomi1998,
  Saxena2000,Tateiwa2001,Settai2002,Pfleiderer2002}
However, further efforts are necessary to understand the experimental results
based on the present theory.
For example, we should calculate the electrical resistivity
to discuss directly the effective mass deduced from $A$ coefficient,
since we cannot resolve the spin components of the effective mass from $A$.

\subsection{Magnetic field effect}
Now, we discuss the magnetic field effect.
We choose $n=1.75$, $V/W=0.2$, and $\epsilon_f/W=0.3$,
which are indicated by the circle in Fig.~\ref{PD}(c).
For this parameter set, the system is paramagnetic
without a magnetic field, but near the ferromagnetic phase boundary.
The effective mass is not large
for this parameter set [see Fig.~\ref{vary_ef_1}(c)].
If we assume a paramagnetic state with a much deeper $f$-level,
we can obtain a large effective mass,
but such a paramagnetic state is unstable against magnetic order
due to the large Coulomb interaction $U$.
Thus, we have chosen the above parameter set.
We believe that the qualitative aspects of $f$-electron systems under a magnetic
field are still captured by the present simple model
with $U \rightarrow \infty$.

\begin{figure}
  \includegraphics[width=0.95\linewidth]
  {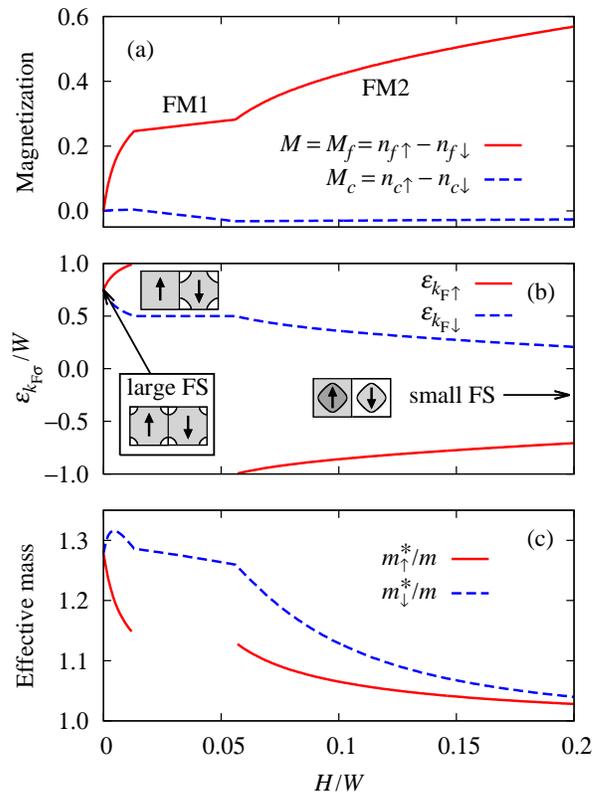}
  \caption{\label{vary_H_1}
    (Color online)
    Physical quantities as functions of $H$
    for $n=1.75$, $V/W=0.2$, and $\epsilon_f/W=0.3$.
    (a) magnetization $M=M_f$ and polarization of the conduction band $M_c$,
    (b) kinetic energy $\epsilon_{k_{\text{F}\sigma}}$
    of the conduction electron at the Fermi momentum,
    and
    (c) effective mass $m^*_{\sigma}$.
    In (b), we show the Fermi-surface structure
    in each phase schematically.
  }
\end{figure}

\begin{figure}
  \includegraphics[width=0.95\linewidth]
  {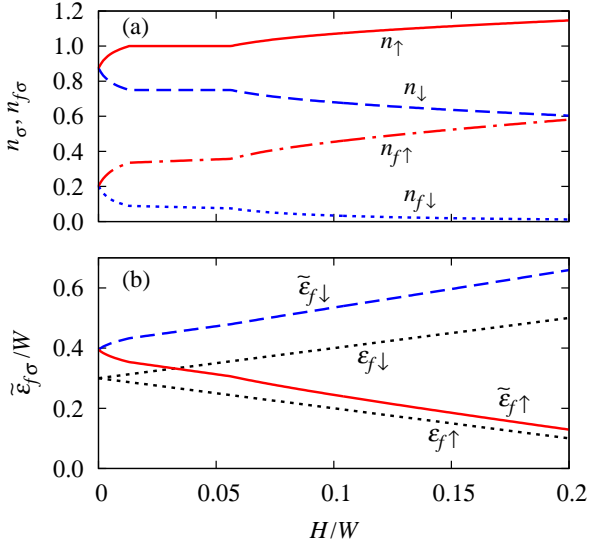}
  \caption{\label{vary_H_2}
    (Color online)
    Physical quantities as functions of $H$
    for $n=1.75$, $V/W=0.2$, and $\epsilon_f/W=0.3$.
    (a) $n_{\sigma}$ and $n_{f\sigma}$
    and
    (b) $\tilde{\epsilon}_{f\sigma}$.
    The dotted lines in (b) indicate the bare $f$-levels
    $\epsilon_{f \sigma}=\epsilon_f-\sigma H$.
  }
\end{figure}

Figure~\ref{vary_H_1} shows the $H$ dependences of the magnetization,
the kinetic energy $\epsilon_{k_{\text{F} \sigma}}$ of the conduction electron
at the Fermi momentum,
and the effective mass.
The polarization of the conduction band $M_c$ is always small
as in the ferromagnetic phases without magnetic field.
The magnetization $M=M_f$ increases continuously as a function of $H$.
The magnetization curve is similar to that in YbRh$_2$Si$_2$,~\cite{Tokiwa2005}
if we regard the anomaly around 10~T in YbRh$_2$Si$_2$ at ambient pressure
as the transition to FM1.
The Fermi-surface structure changes continuously
from the large Fermi-surface in PM
to the small Fermi-surface in FM2.
The effective mass decreases by a magnetic field except for a small-$H$ region,
since a magnetic field polarizes the $f$ electrons
and the correlation effect becomes weak.
In the small-$H$ region,
$n_{f \uparrow}$ increases by $H$ but $n_{f \downarrow}$ is not very small
[see Fig.~\ref{vary_H_2}(a)],
and the effect of the Coulomb interaction becomes stronger
for the down-spin state.
Then, the effective mass for the down-spin electrons
increases as $H$ in the small-$H$ region, and has a peak.

Figure~\ref{vary_H_2} shows the magnetic field dependences of
$n_{\sigma}$,  $n_{f \sigma}$, and $\tilde{\epsilon}_{f \sigma}$.
By increasing $H$, the system turns into FM1 with $n_{\uparrow}=1$.
By increasing $H$ further, the system turns into FM2,
and the polarization of $f$ electrons approaches
the saturation value asymptotically, i.e.,
$n_{f \uparrow} \rightarrow 1$ and $n_{f \downarrow} \rightarrow 0$.
The renormalized $f$-level changes monotonically.
$\tilde{\epsilon}_{f \uparrow}$ becomes very close to $\epsilon_{f \uparrow}$
by increasing $H$,
since the correlation effects on the up-spin electrons are weak
for $n_{f \downarrow} \rightarrow 0$.

There are kinks in all the above quantities at the transition points
to FM1 and from FM1 to FM2.
The kinks in $\tilde{\epsilon}_{f \sigma}$ [Fig.~\ref{vary_H_2}(b)]
are weak and invisible on this scale.
While these ferromagnetic transitions are continuous,
they are not crossovers even under magnetic fields.
We discuss this issue in the next subsection.

\subsection{Order of the ferromagnetic phase transitions}
\begin{figure}
  \includegraphics[width=0.95\linewidth]
  {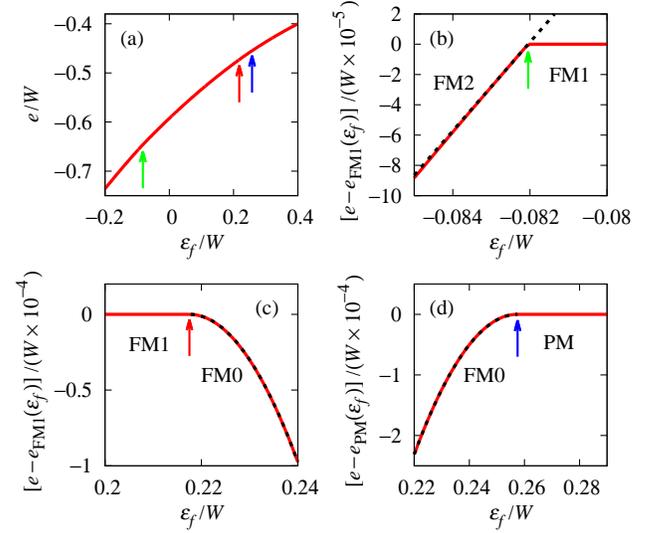}
  \caption{\label{energy}
    (Color online)
    Energy $e$ per site as a function of $\epsilon_f$
    for $n=1.75$ and $V/W=0.2$.
    (a) energy in a wide range of $\epsilon_f$.
    Arrows indicate the phase transition points.
    (b)-(d) show energy around each phase transition point:
    (b) between FM2 and FM1,
    (c) between FM1 and FM0,
    and
    (d) between FM0 and PM.
    Energy in phase X is fitted by a polynomial function
    $e_{\text{X}}(\epsilon_f)$ and we subtracted $e_{\text{X}}(\epsilon_f)$
    from $e$ in (b)-(d).
    This difference is linear in $\epsilon_f$ in (b)
    and is quadratic in $\epsilon_f$ in (c) and (d)
    as indicated by the dotted lines.
  }
\end{figure}

In this subsection, we discuss the order of the phase transitions.
It is usual that between ferromagnetic states,
the transition is a first-order phase transition
or just a crossover, not a phase transition,
since the symmetry is the same between the ferromagnetic states.
However, in the present model,
the transitions between the ferromagnetic phases, FM0, FM1, and FM2,
can be phase transitions even if they are continuous.
To explicitly demonstrate it,
we show the energy $e$ per site in Fig.~\ref{energy}
as a function of $\epsilon_f$ for $n=1.75$ and $V/W=0.2$
without a magnetic field, around the phase transition points.
The first derivative of $e$ has a jump at the FM2-FM1 boundary
as shown in Fig.~\ref{energy}(b),
and it is a first-order phase transition.
The second derivative of $e$ has a jump at the FM1-FM0 boundary
as shown in Figs.~\ref{energy} (c),
and it is a second-order phase transition, not a crossover.
The FM0-PM phase transition is also of second order.

We can show that when the magnetization $M$ changes its slope
but is continuous at a point,
as in Figs.~\ref{vary_ef_1}(a) and \ref{vary_H_1}(a),
it is a second-order phase transition point (see Appendix).
That is, to cause a second-order phase transition,
it is not necessary to break symmetry.
Behind such a second-order phase transition between ferromagnetic states
in the present model,
the topology of the Fermi surface changes, i.e., it is a Lifshitz transition.
Note that while the originally proposed Lifshitz transition
is of 2.5 order,~\cite{Lifshitz1960}
the present Lifshitz transition accompanying magnetism is of second order.

Note also that the transitions to FM1
and from FM1 to FM2 under magnetic fields
shown in Figs.~\ref{vary_H_1} and \ref{vary_H_2}
are second-order phase transitions,
while, in ordinary cases, a continuous ferromagnetic transition
becomes a crossover under a finite magnetic field.
These transitions under magnetic fields are possible even for $U=0$.

At finite temperatures,
the second-order phase transitions between ferromagnetic phases
would become crossovers,
since the Fermi surface is not well defined at finite temperatures.
On the other hand, first-order phase transitions are possible
even at finite temperatures.

\section{Antiferromagnetic states}\label{AF}
In the present study,
we have assumed uniform states: paramagnetic and ferromagnetic.
The calculated results have been interpreted with the aid of
the schematic bands shown in Fig.~\ref{bands}.
A similar discussion may be applicable to antiferromagnetic states.

\begin{figure}
  \includegraphics[width=0.98\linewidth]
  {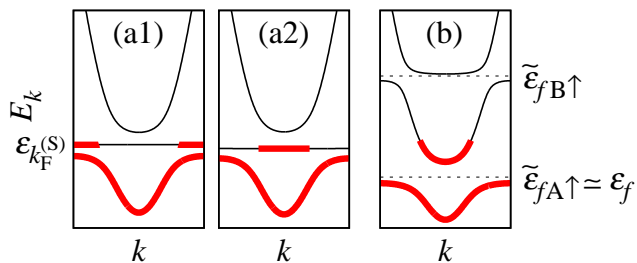}
  \caption{\label{AFbands}
    (Color online)
    Schematic band structures expected in antiferromagnetic phases.
    In a weakly polarized state,
    the band structure like (a1) or (a2) will realize.
    In a strongly polarized state,
    the band structure shown in (b) will realize.
    $E_k$ denotes the quasi-particle energy.
    The bold lines indicate the states occupied by electrons.
  }
\end{figure}

We show the schematic bands
expected in antiferromagnetic states with a two-sublattice structure
in Fig.~\ref{AFbands}.
In a weak antiferromagnetic state,
the difference of the renormalized $f$-level between
A and B sublattices is small,
and we may obtain the band structure by simply folding the Brillouin zone
as shown in Fig.~\ref{AFbands}(a1).
In a strongly polarized antiferromagnetic state,
the renormalized $f$-levels will be much different between A and B sublattices.
We show a schematic band structure in such a state in Fig.~\ref{AFbands}(b),
by assuming that the $f$ orbitals on the A sublattice are mainly occupied
by up-spin electrons
and the $f$ orbitals on the B sublattice are mainly occupied
by down-spin electrons.
The effective $f$-level $\tilde{\epsilon}_{f\text{A}\uparrow}$
for up spin on the A sublattice is almost the same
as the bare $f$-level $\epsilon_f$
and the effective $f$-level $\tilde{\epsilon}_{f\text{B}\uparrow}$
for up spin on the B sublattice is much higher than the Fermi level
[cf. the ferromagnetic case Fig.~\ref{bands}(d)].
Note that
$\tilde{\epsilon}_{f\text{A}\uparrow}=\tilde{\epsilon}_{f\text{B}\downarrow}$
and
$\tilde{\epsilon}_{f\text{B}\uparrow}=\tilde{\epsilon}_{f\text{A}\downarrow}$.
In the strongly polarized state,
the electronic state around the Fermi surface is mainly composed of
the conduction-electron states.

Since the topology of the Fermi surfaces are different between (a1) and (b),
a phase transition takes place as the antiferromagnetic moment develops
provided the system first turns into the antiferromagnetic state
with the band structure (a1) from the paramagnetic state.
Indeed, such a phase transition in the antiferromagnetic phase has been found
in the Kondo lattice model~\cite{Watanabe2007,Lanata2008,Martin2010}
and in the periodic Anderson model,~\cite{Watanabe2009}
and possibility to explain the Fermi-surface reconstruction
in CeRh$_{1-x}$Co$_x$In$_5$~\cite{Goh2008}
and in YbRh$_2$Si$_2$ with chemical pressure~\cite{Friedemann2009}
has been discussed.

In addition, the direct transition from the paramagnetic state
to the antiferromagnetic state with the band structure
shown in Fig.~\ref{AFbands}(a2),
which has the same topology of the Fermi surface as in (b), is possible,
since the band originates from the $f$ orbital is very flat.
Note that the band structure (a2) is not obtained by simply folding
that in the paramagnetic state,
and the effects of the change in the Fermi surface would be drastic.
This transition has also been found
in the Kondo lattice model~\cite{Watanabe2007,Lanata2008}
and in the periodic Anderson model,~\cite{Watanabe2009}
and has been proposed as a possible mechanism to explain the change
in the Hall coefficient of YbRh$_2$Si$_2$
at the antiferromagnetic quantum critical point.~\cite{Paschen2004}
Note that we expect a mass enhancement around such
a magnetic transition point as we have shown for the ferromagnetic case.
Thus, this transition may also be a candidate for the mechanism of
the Fermi-surface change and the enhancement of the effective mass
around the antiferromagnetic transition point
observed by the de Haas-van Alphen experiments under pressure
on CeRh$_2$Si$_2$,~\cite{Araki2001}
CeRhIn$_5$,~\cite{Shishido2005}
and CeIn$_3$.~\cite{Settai2005}

\section{Summary}\label{summary}
We have studied the ferromagnetism and the magnetic field effect
in the periodic Anderson model by using the Gutzwiller theory.
There are three ferromagnetic phases, FM0, FM1, and FM2.
The Fermi-surface structure changes according to the magnetic state.
The PM state has a large Fermi-surface,
the FM0 state is a weak ferromagnetic state,
the FM1 state is a half-metallic state
without a Fermi surface for up-spin electrons,
and the FM2 state has a small Fermi-surface.
The effective mass has a peak in the FM1 phase as a function of $\epsilon_f$.

The transitions between these ferromagnetic phases
can be second-order phase transitions,
while we cannot define the order parameter in an ordinary way
due to the absence of symmetry breaking.
These second-order phase transitions originate from
the change in the Fermi-surface topology
and are called Lifshitz transitions.
We have found that the present Lifshitz transitions
accompanying magnetism are of second order,
while the originally proposed Lifshitz transition
is of 2.5 order.~\cite{Lifshitz1960}

According to the theory of phase transitions,
if the symmetry is broken spontaneously,
a phase transition takes place.
However, the converse is not necessarily true.
For example, the liquid-vapor transition of water
is a first-order phase transition without symmetry breaking.
In the present paper,
we have shown that a second-order transition is also possible
without symmetry breaking.

In the present model with $U \rightarrow \infty$,
a paramagnetic state with a large mass enhancement is not attained,
since the magnetically ordered state becomes stable
against the paramagnetic state before the effective mass becomes very large.
Thus, we should revise the present model
to describe the heavy-fermion state quantitatively, e.g.,
by using a finite value of $U$ and/or
by introducing the orbital degrees of freedom of $f$ electrons.~\cite{Rice1986}
It is an important future problem.

\begin{acknowledgments}
  This work is supported by
  a Grant-in-Aid for Young Scientists (B) from
  the Japan Society for the Promotion of Science.
\end{acknowledgments}

\appendix*
\section{Sufficient condition for a second-order phase transition}
A second-order phase transition is defined by a jump in the second derivative
of the free energy (or energy at zero temperature).
We consider the system described by the free energy $F(x,M)$.
$x$ is a controlling parameter such as magnetic field,
pressure, $f$-electron level, and temperature.
$M$ represents a physical quantity
such as magnetization and $f$-electron number.
The physical quantity $M(x)$ at $x$ is determined
by minimizing $F(x,M)$ with respect to $M$:
\begin{equation}
  \left.\frac{\partial F(x,M)}{\partial M}\right|_{M=M(x)}=0.
\end{equation}
Then, the first derivative of the free energy $F(x,M(x))$ at $x$ is
\begin{equation}
  \frac{d F(x,M(x))}{dx}
  =\left.\frac{\partial F(x,M)}{\partial x}\right|_{M=M(x)}.
\end{equation}
If $M(x)$ changes discontinuously at a point,
the first derivative has a jump at this point
and it is a first-order phase transition.
The second derivative is given by
\begin{equation}
  \begin{split}
    \frac{d^2 F(x,M(x))}{dx^2}
    =&\left.\frac{\partial^2 F(x,M)}{\partial x^2}\right|_{M=M(x)}\\
    +&\left.\frac{\partial^2 F(x,M)}{\partial x \partial M}\right|_{M=M(x)}
    \frac{dM(x)}{dx}.
  \end{split}
\end{equation}
Then, if $M(x)$ is continuous and $dM(x)/dx$ is discontinuous at a point,
it is a second-order phase transition.

We have not assumed that $M(x)=0$ below or above the transition point.
Thus, the above discussion does not require
that $M$ is an order parameter to describe symmetry breaking.


%

\end{document}